\documentclass[referee]{raa}
\usepackage{txfonts}
\usepackage{natbib}  % added by zhijia
\usepackage{graphicx,times}
\usepackage{amssymb}
\usepackage{threeparttable}
\usepackage{longtable}
\usepackage{color}
\usepackage{setspace}

\begin{document}
\large{}

   \title{Variations of helioseismic parameters due to magnetic field generated by a flux transport model}

    \volnopage{Vol.0 (200x) No.0, 000--000}      %%preserved for Editor. DOn't remove!
   \setcounter{page}{1}          %%starting page, preserved for Editor. DOn't remove!
  % \titlerunning{Helioseismic Diagnostics of solar magnetic fields}
   %\subtitle{Helioseismic diagnostics of solar magnetic fields}

   \author{Shaolan Bi
          \inst{1}
          \and
          Tanda Li%\inst{2}\fnmsep\thanks{Just to show the usage
%          of the elements in the author field}
          \inst{2,5}
          \and
          Kang Liu
          \inst{1}
          \and
          Jie Jiang
          \inst{3}
          \and
          Yaguang Li
          \inst{4,5}
          \and
          Jinghua Zhang
          \inst{6,1}
          \and
          Xianfei Zhang
          \inst{1}
           \and
          Yaqian Wu
          \inst{6}
          }

   \institute{Department of Astronomy, Beijing Normal University,
             Beijing 100875, China; {\it bisl@bnu.edu.cn; liukang@bnu.edu.cn}\\
             \and
             School of Physics and Astronomy, University of Birmingham, Edgbaston, Birmingham, B15 2TT, UK; {\it t.li.2@bham.ac.uk}\\
             \and
             School of Space and Environment, Beihang University,
             Beijing 100871, China\\
             \and
             Sydney Institute for Astronomy (SIfA), School of Physics, University of Sydney, NSW 2006, Australia\\
             \and
             Stellar Astrophysics Centre, Department of Physics and Astronomy, Aarhus University, Ny Munkegade 120, DK-8000 Aarhus C, Denmark\\
             \and
             National Astronomical Observatories, Chinese Academy of Science,
             Beijing 100012, China\\
             }

  % \date{Received September 15, 1996; accepted March 16, 1997}
 \abstract{The change of sound speed has been found at the base of the convection during the
   solar cycles, which can be used to constrain the solar internal magnetic field.
   We aim to check whether the magnetic field generated by the solar dynamo
   can lead to the cyclic variation of the sound speed detected through helioseismology.
   The basic configuration of magnetic field in the solar interior was obtained
   by using a Babcock-Leighton (BL) type flux transport dynamo. We reconstructed one-dimensional
   solar models by assimilating magnetic field generated by an established dynamo and examined their influences on the structural variables. The results show that  magnetic field generated by the dynamo is able to cause noticeable change of the sound speed profile at the base of the convective zone during a solar cycle. Detailed features of this theoretical prediction are also similar to those of the helioseismic results in solar cycle 23  by adjusting the free parameters of the dynamo model.
 \keywords{Sun: oscillations --
             Sun: activity --
             Sun: interior
               }
}

%   \authorrunning{Helioseismic Diagnostics of Cycle Magnetic Fields of the Solar Interior}
%   \titlerunning{Bi et al}     %\authorrunning{Bi et al.}
 \maketitle

%
%________________________________________________________________

\section{Introduction}

  Helioseimology has been regarded as a powerful tool to detect the properties
  of the solar interior that are not directly observed \citep{christensen-Dalsgaard}.
  Global helioseimology utilizes the normal modes of oscillation of the Sun to determine
  the interior structure and dynamics. The oscillation frequencies are known to vary
  on timescales related to the solar cycles
  \citep[e.g.,][]{woodard,libbrecht90,elsworth90,basu00,howe00}.
  The change of frequency has been shown to be highly correlated with surface activity
  \citep[e.g.,][]{chaplin07,jain09,broomhall09,tripathy15,howe18}. They concluded that
  the observed frequency change is confined to the shallow layer of the Sun.
  In addition, \citet{howe02} showed that the temporal and latitudinal distribution
  of the frequency shifts is correlated with the distribution of the surface magnetic
  field.

  With improved data and analysis techniques in recent years, helioseimology
  has successfully probed the structural changes in the deeper layers of
  the convective zone, especially the tachocline at the base of the convective zone.
  Although the solar oscillation frequencies have been determined with tremendous
  precision, statistical errors in those frequencies are still too large to make
  any direct detections of structural change in the deep interior.
  Two major approaches  were suggested to meet these challenges.
  One is to use the smoothed and scaled frequency change as a function of the lower turning
  point \citep{chou05,serebryanskiy05}. The other one is to use a principal
  component analysis (PCA) method to separate the frequency differences into a linear
  combination of different time-dependent components \citep{baldner08}.
  In both cases, a small but statistically significant change in the sound speed
  with an origin at and below the base of the convection zone was found. By assuming
  that the entire change is due to the presence of magnetic field, they constrained
  a magnetic field strength in the order of $\sim10^5$G. \citet{baldner08}
  and \citet{baldner09} also showed that the sound speed inversions are tightly correlated
  with the latitudinal distribution of surface activity. Besides,
  \citet{liang15} presented the travel time difference which was attributed to the change of magnetic field. Therefore, combining the observed
  sound speed variation with the frequency shift would provide more constraints on the
  configuration of the magnetic field deep inside the Sun.

  It is widely accepted that all the solar activities are dominated by the solar magnetic
  field generated inside the Sun due to the dynamo process. Where the magnetic field is
  generated and how the magnetic field is distributed are longstanding and outstanding questions
  in solar physics. Up to date, a number of dynamo models have been developed for investigating
  the dynamo process. The details of models can be found in the reviews by \citet{charbonneau,charbonneau14}.
  Global MHD simulation of the solar convective zone is the most direct way to
  tackle the solar convective zone. Simulations of convection-driven dynamos have recently reached
  a level of sophistication \citep{hotta2016,strugarek2017}. However, due to a
  wide range of spatial and temporal scales characterizing the solar convection,
  the variability seen in the simulations is not directly comparable to that of the Sun.
  The kinematic flux transport dynamo (FTD) model based on the Babcock-Leighton (BL)
  mechanism, which was first proposed by \citet{babcock61} and further elaborated
  by \citet{leighton64}, is regarded as one of the most
  promising models in understanding the solar cycle during the past several years
  \citep[e.g.,][]{jiang07,cameron10,jiang13}. Thanks to the fundamental works of
  \citet{nandy02} and \citet{chatterjee04}, the code SURYA based on the
  FTD model has been well developed and open to the public for years \citep{Choudhuri2017}.

 With the variable magnetic field and turbulence included, one-dimensional models of the structure and evolution of the Sun were constructed, which were then compared to observations
  \citep{li03}. Since the magnetic configuration is unknown, \citet{li03}
  assumed a Gaussian profile of the magnetic field concentrated at different depths with different amplitudes.
  They found a model with magnetically modulated turbulence which reproduces shifts of oscillation frequencies observed in the solar cycle 23.
  This result, however, contains an obvious limitation. That is the simple descriptions of magnetic field during a solar cycle: a Gaussian distribution below the surface with varying amplitude.
 Therefore, we aim to develop the solar variability model field, e.g., those generated by the FTD models. Different from \cite{li03}, this work will focus on the structural variations at the tachocline where strong magnetic field is generated.

 In this work, we adopted the code SURYA to generate generate a series of magnetic profiles through a complete solar cycle, and then incorporate the self-consistent magnetic fields into the computation of stellar evolution models for investigating the effects of the magnetic fields on the structural properties and the oscillation frequencies. In Section 2, we describe the physical ingredients of the dynamo model and the solar variability model. Section 3 presents the details of the magnetic profiles generated by a FTD model in a complete solar cycle. In Section 4, we show  the impacts on the solar internal structural variables due to magnetic field. The discussions and conclusions are given in Section 5.

\section{Theoretical Models}

  In this section, we briefly introduce the physical ingredients for a
  BL type flux transport dynamo model, and for a solar variability
  model that includes the effects of magnetic field and rotation.

\subsection{The Flux Transport Dynamo Model}

  Solar magnetic activity involves the generation and evolution of magnetic field.
  The important ingredients in the flux transport dynamo model are as follows.
  (1) The strong toroidal field is produced by stretching of the poloidal field
  lines, which is caused by the differential rotation within the tachocline where
  the rotational velocity sharply changes with depth and latitude; (2) When the toroidal field
  $B_{\rm t}$ exceeds the critical field value $B_{\rm c}$, the tachocline toroidal
  field undergo buoyant rise through the convection zone to produce sunspots;
  (3) the poloidal field can be generated by the BL process;
  (4) The meridional circulation plays an important role for the advection of the
  toroidal and poloidal field \citep{chatterjee04,jiang07,Choudhuri2020}.

  In the spherical polar coordinates $\left(r,\theta,\phi\right)$, the averaged
  large-scale magnetic field and plasma flow under the assumption of axisymmetry
  about the Sun's rotation axis, can be expressed as
  \begin{equation}
  \mathbf{B}=B_{\rm t}\left(r,\theta\right)\mathbf{e}_{\phi}
  +\nabla\times\left[A\left(r,\theta\right)\bf{e}_{\phi}\right],
  \end{equation}
  \begin{equation}
  \mathbf{v}=\Omega\left(r,\theta\right)r\sin\theta\mathbf{e}_{\phi}
  +\mathbf{v}_{\rm p},
  \end{equation}
  where $\mathbf{B}_{\rm t}=B_{\rm t}\left(r,\theta\right)\mathbf{e}_{\phi}$,
  $\mathbf{B}_{\rm p}=\nabla\times\left[A\left(r,\theta\right)\mathbf{e}_{\phi}\right]$
  are the toroidal field and poloidal field, respectively.
  The first term of equation (2) denotes the $\phi$-component
  of the velocity, i.e., angular velocity $\Omega(r,\theta)$ of the solar interior
  inferred from helioseismic data \citep{kosovichev,schou98}, while
  $\mathbf{v}_{\rm p}={\rm v}_{\rm r}(r,\theta)\mathbf{e}_{\rm r}+
  {\rm v}_{\theta}(r,\theta)\textbf{e}_{\theta}$
  is the meridional circulation. The equations for the standard $\alpha\Omega$
  dynamo model are given as follows:
  \begin{equation}
  \frac{\partial A}{\partial
  t}=\eta_{\rm p}\left(\nabla^{2}-\frac{1}{s^{2}}\right)A
  -\frac{1}{s}\left(\mathbf{v}_{\rm p}\cdot\nabla\right)\left(sA\right)
  +\alpha B_{\rm t},
  \end{equation}
  \begin{eqnarray}
  \frac{\partial B_{\rm t}}{\partial t} & = &
  \eta_{\rm t}\left(\nabla^{2}-\frac{1}{s^{2}}\right)B_{\rm t}
  +\frac{1}{r}\frac{d\eta_{\rm t}}{dr}\frac{\partial}{\partial
  r}\left(rB_{\rm t}\right)
  \\ & - & \frac{1}{r} \left[\frac{\partial}{\partial
  r}\left(r{\rm v}_{\rm r}B_{\rm t}\right)+ \frac{\partial}{\partial
  \theta}\left({\rm v}_{\theta}B_{\rm t}\right)\right]
  +s\left(\mathbf{B}_{\rm p}\cdot\nabla\right)\Omega, \nonumber
  \end{eqnarray}
  where $s=r\sin\theta$. The turbulence diffusion coefficients $\eta_{\rm p}$
  and $\eta_{\rm t}$ correspond to the poloidal and toroidal components,
  respectively. The coefficient $\alpha$ expresses a BL source term which
  describes the generation of poloidal field due to the buoyant eruption
  and flux dispersal of tilted active regions.

  Here we describe a few key parameters in particular. The meridional flow  plays essential roles
  in the BL type dynamo. It dominates the cycle period and  is mainly responsible for
  the equatorward migration of the toroidal field and poleward migration of
  the poloidal field on the solar surface. It is noted that the penetration depth and the number of circulation cells are still the subjects of hot debate.
  We follow \citet{chatterjee04} to adopt a deep penetrated one-cell meridional flow.
  It goes slightly below the tachocline until $0.61R_{\odot}$. A strong turbulent diffusivity
  $\eta_p=2.6\times10^{12}$ cm$^2$ s$^{-1}$ is adopted for the poloidal field, which corresponds
  to the diffusion-dominated flux transport dynamos. This distinguishes from the advection-dominated
  ones with low turbulent diffusivity. The different strength of $\eta_p$
  has large effects on the path of the flux transport and flux structure in the convective zone.
  The $\alpha$-effect is concentrated in the top layer $0.95R_{\odot}\leq r \leq R_{\odot}$
  , where $\alpha$ changes with latitude as $\cos\theta$. The only nonlinear suppression of the magnetic
  field growth is provided by magnetic buoyancy. The magnetic buoyancy is dealt in the same way as
  \citet{chatterjee04}. A critical field $B_c$ is set. Wherever the toroidal field $B$ exceeds $B_c$,
  a fraction $f$ = 0.5 of the magnetic flux is assumed to erupt to the surface layers, with the toroidal field values
  adjusted appropriately to ensure flux conservation. The rest part of the dynamo system is linear.
  The adopted value of $B_c$ sets the magnetic field scale of the solutions. It will be an adjustable
  parameter in Section \ref{sec:SUYRA} to make a constraint on the possible field strength in
  the convective zone. The widely studied parameters, such as turbulent pumping, are not included in the model \citep{jiang13}.

  The axisymmetric dynamo equations (3) and (4) are to be solved in a
  meridional slab, i.e., $R_{\rm b}\leq r \leq R_{\odot}$ and $0\leq \theta \leq \pi$,
  with the inner boundary at $R_{\rm b}=0.55R_{\odot}$.
  Assuming a perfectly conducting solar core, at the inner radius
  $(r = R_{b})$, or at the poles $(\theta=0,\pi)$, we have
  \begin{equation}
  A=0, \ \ \
  B_t=0.
  \end{equation}
  In general, it is assumed that the Sun is in a vacuum without
  electrical currents, i.e., $\nabla\times\textbf{B}=0$. At the top $\left(r=R_{\odot}\right)$,
  the toroidal field has to be zero and the poloidal field has to
  match smoothly a potential field satisfying the free space equation, this requires
  \begin{equation}
  \left(\nabla^{2}-\frac{1}{r^{2}\sin^{2}\theta}\right)A=0, \ \ \
  B_t=0.
  \end{equation}
  The wider radial region of our calculated spherical shell compared to other models,
  which are usually in the range of $\left(0.65R_{\odot}, 1.0R_{\odot}\right)$, makes the comparisons with
  the helioseismologic results more feasible.

\subsection{1-D Solar Models with Magnetic field}

  When the influence of a cyclic magnetic field is considered,
  the solar models become variable \citep[see][]{li03}. Cyclic magnetic field $\mathbf{B}$,
  given by the flux transport dynamo model, is a vector with two components.
  Within the framework of 1-D stellar evolution model, instead of using the
  two components, new variables were introduced to the stellar
  basic equations, namely, the magnetic energy per unit mass $\chi$ and magnetic field
  direction $\gamma$, which are defined as \citep{lydon95}:
  \begin{equation}
  \chi = (B^{2}/8\pi)/\rho , \ \ \
  \gamma=1+B^{2}_{\rm t}/B^{2},
  \end{equation}
  where $B^{2}=B^{2}_{\rm t}+B^{2}_{\rm p}$.

  Following \citet{li01} and \citet{li03}, the stellar structure variables,
  $\chi$ and $\gamma$, can be used to describe the magnetic structure of a star
  in the one-dimensional stellar modeling.
  The magnetic pressure $P_{\rm m}$ can be defined as:
  \begin{equation}
  P_{\rm m}=(\gamma-1)\chi\rho.
  \end{equation}
  The equation of state is modified as $\rho=\rho\left(P,T, \chi, \gamma\right)$,
  and the corresponding differential form is given by
  \begin{equation}
  \frac{d\rho}{\rho}=\zeta\frac{dP}{P}-\delta\frac{dT}{T}-\lambda\frac{d\chi}{\chi}
  -\mu\frac{d\gamma}{\gamma}.
  \end{equation}
  The first law of thermodynamics should be written as
  \begin{eqnarray}
  TdS & = & dU+PdV-d\chi, \\
      & = & c_{P}dT-\frac{\delta}{\rho}dP
      +\left(\frac{P\delta\lambda}{\rho\zeta\chi}-1\right)d\chi
  +\frac{P\delta\mu}{\rho\zeta\gamma}d\gamma, \nonumber
  \end{eqnarray}
  where the total pressure is defined as $P=P_{0}+P_{\rm m}$, and $P_{0}$
  is the gas pressure. The related derivatives are
  \begin{displaymath}
  \zeta = \left(\frac{\partial \ln\rho}{\partial \ln
  P}\right)_{\rm T,\chi,\gamma} , \ \ \delta=-\left(\frac{\partial
  \ln\rho}{\partial \ln T}\right)_{\rm P,\chi,\gamma} ,
  \end{displaymath}
  \begin{displaymath}
  \lambda  =-
   \left(\frac{\partial \ln\rho}{\partial \ln \chi}\right)_{\rm P,\rm T,\gamma}
   ,\ \
  \mu=-\left(\frac{\partial \ln\rho}{\partial \ln
  \gamma}\right)_{\rm P,\rm T,\chi} .
  \end{displaymath}
  A detailed derivation of the solar variable model, which includes the effect of
  magnetic field, was described in \citet{lydon95} and \citet{li01}.

  Consequently, when magnetic field and rotation are included, the
  stellar structure equations are modified by the following
  \citep{denissenkov07,eggenberger08}
  \begin{eqnarray}
  \frac{\partial P}{\partial M_{\rm P}} & = & -\frac{GM_{\rm P}}{4\pi
  r_{\rm P}^{4}}f_{\rm P}, \\
  \frac{\partial r_{\rm P}}{\partial M_{\rm P}} & = & \frac{1}{4\pi
  r_{\rm P}^{2}\rho}, \\
  \frac{\partial L_{\rm P}}{\partial M_{\rm P}} & = & \epsilon-T\frac{dS}{dt},
  \\
  \frac{\partial T}{\partial M_{\rm P}} & = & -\frac{GM_{\rm P}}{4\pi
  r_{\rm p}^{4}}f_{\rm P}\min\left[\nabla_{\rm con},
  \nabla_{\rm rad}\frac{f_{\rm T}}{f_{\rm P}}\right],
  \end{eqnarray}
  where the subscript $P$ refers to the isobar value. The nondimensional rotating corrective
  factors $f_{\rm P}$ and $f_{\rm T}$ depend on the shape of the isobars, namely
  \begin{displaymath}
  f_{\rm P}=-\frac{4\pi r_{\rm p}^{4}}{GM_{\rm P}S_{\rm P}}\frac{1}{\langle
  g^{-1}\rangle} , \\
  \end{displaymath}
  \begin{displaymath}
  f_{\rm T}=\left(\frac{4\pi r_{\rm p}^{2}}{S_{\rm P}}\right)^{2}\frac{1}{\langle
  g\rangle\langle g^{-1}\rangle}.
  \end{displaymath}
  Here $\langle g \rangle$ and $\langle g^{-1} \rangle$ are the mean values
  of the effective gravity and its inverse over the equipotential surface.
  $S_{\rm p}$ is the surface area of the equipotential, while other variables have
  been described by \citet{meynet97}.

\section{ Magnetic field generated by FTD models} \label{sec:SUYRA}

\begin{figure}
 \centering
 \resizebox{0.8 \hsize}{!}{\includegraphics{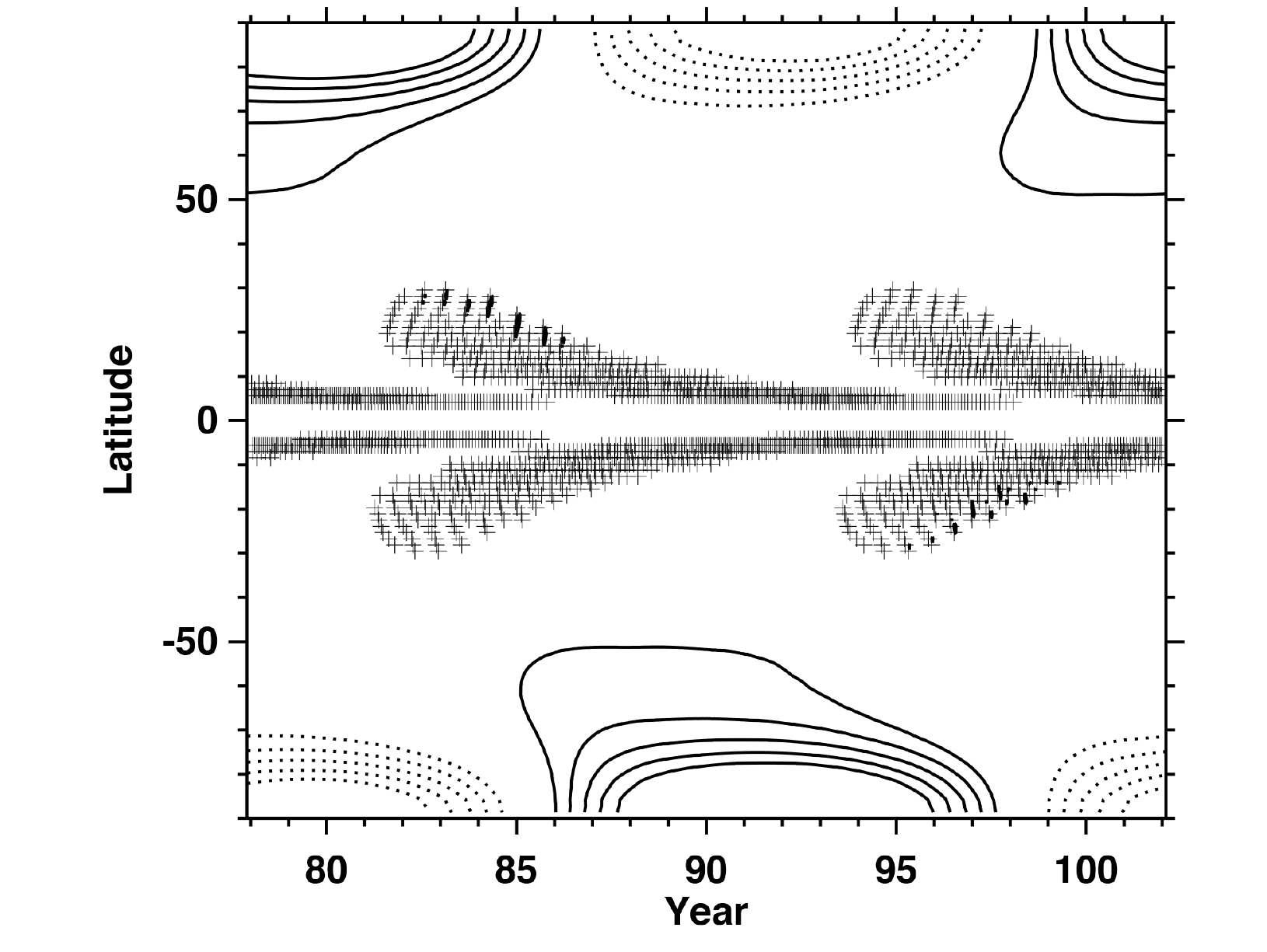}}
 \caption{Theoretical butterfly diagram of eruptions for the simulation given
  by SURYA. The time and latitude of eruption for toroidal field are denoted by crosses.
 The solid and dashed lines are  the  contours of diffuse radial field. The dashed contours
  are for negative $B_{\rm r}$, and the solid contours are for positive $B_{\rm r}$.  }
 \label{Fig:1}
 \end{figure}

In this section, we demonstrate the details of the magnetic field generated by a FTD model (SUYRA) in a complete solar cycle and how we transform magnetic field from 2-D to 1-D for assimilating them into a solar model.
We generated a series of magnetic profiles with SURYA. There are three adjusted input parameters and we set up them following \citet{chatterjee04}. Magnetic buoyancy is prescribed in the way that the toroidal field exceeding the critical field $B_{c}$ is searched above the base of convective zone taken at $r$ = 0.71. Wherever the magnetic strength exceeds $B_{c}$, a fraction of $f$ = 0.5 of it is made to erupt to the surface layers, with the toroidal field values adjusted appropriately to ensure flux conservation. We refer this dynamo model as the SURYA Standard Case in the following analysis.
Figure~\ref{Fig:1} shows the butterfly diagram of eruptions when $B_c$ is equal to $2\times10^5$ G. The solid and dashed lines are  the contours of the radial field. The sunspot eruptions
are confined within $\pm40^{\circ}$ and the butterfly diagrams have shapes similar to  observations. The weak radial field migrates poleward at higher latitudes. The phase relation
between the sunspots and the weak diffuse field is also produced. All of these are consistent
with the observed magnetic butterfly diagram.

\begin{figure}
  \centering
  \resizebox{0.60\hsize}{!}{\includegraphics{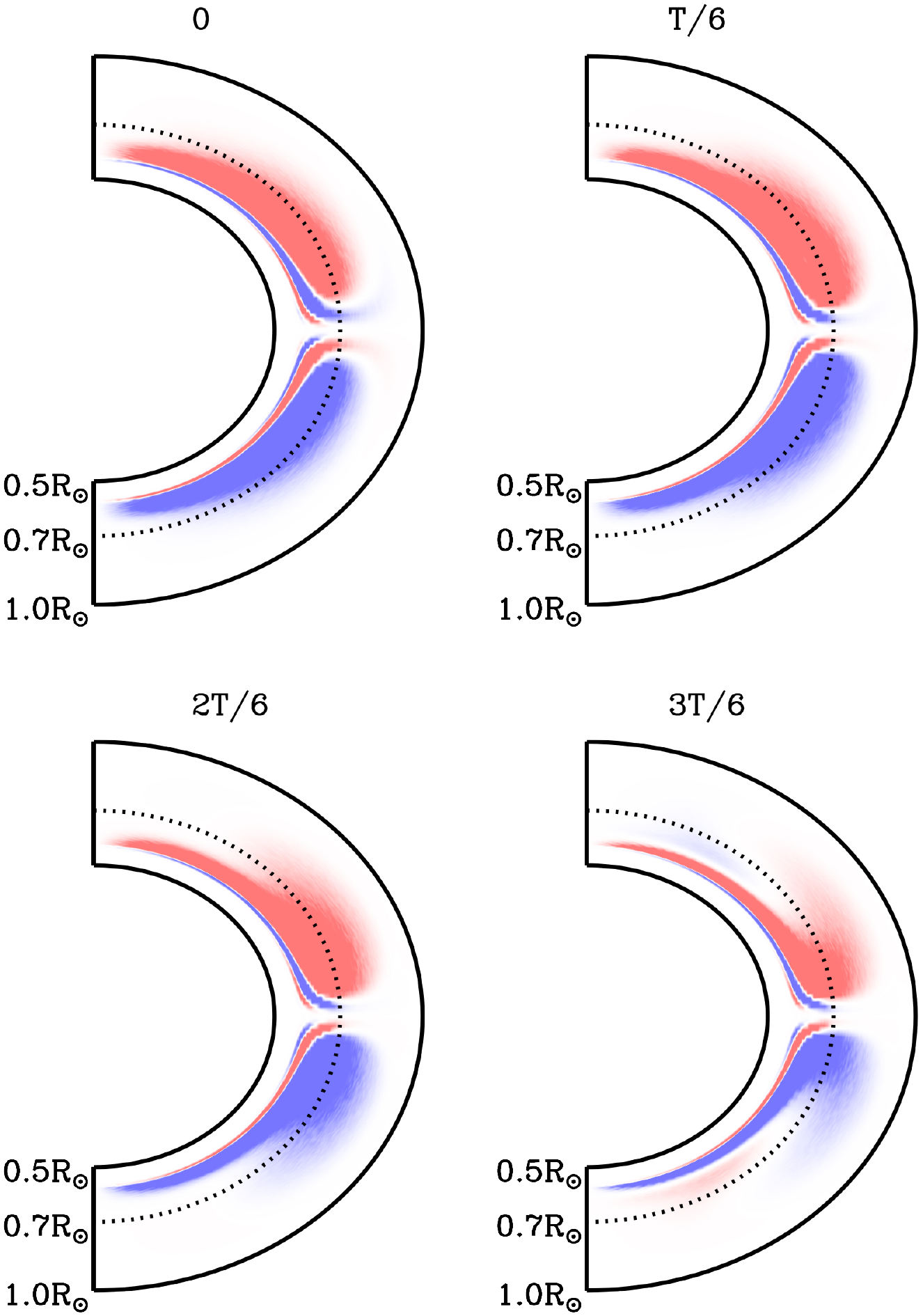}}
  \caption{Theoretical distribution of toroidal magnetic field
  generated by a BL type flux transport dynamo model. Positive
  $B_{\rm t}$ is showed in blue, and negative $B_{\rm t}$ is plotted in
  red. }
  \label{Fig:2}
  \end{figure}

  Figure~\ref{Fig:2} displays the 2-D distribution of the toroidal field in red and blue colors and
  the poloidal field in solid and dashed curves at an interval of 1/6 the solar cycle
  period in the first half of one solar cycle, ordered from the minimum to the maximum.
  As presented in \citet{jiang07}, the poloidal field is radially transported to the bottom
  of the convective zone under the effect of strong turbulent diffusion and is poleward
  transported to the pole due to the effect of poleward meridional flow simultaneously.
  The arrival of the poloidal field changes the strength of the toroidal field.
  The toroidal field distribution is also affected by the transport of the deep
  penetrated meridional flow. The strong effects of the meridional flow and the
  radial shear at the base of the convective zone cause the multi structures
  and evolution of the toroidal field around 0.6-$0.7R_{\odot}$.

\begin{figure}
  \centering
  \includegraphics[width=120mm, angle=0]{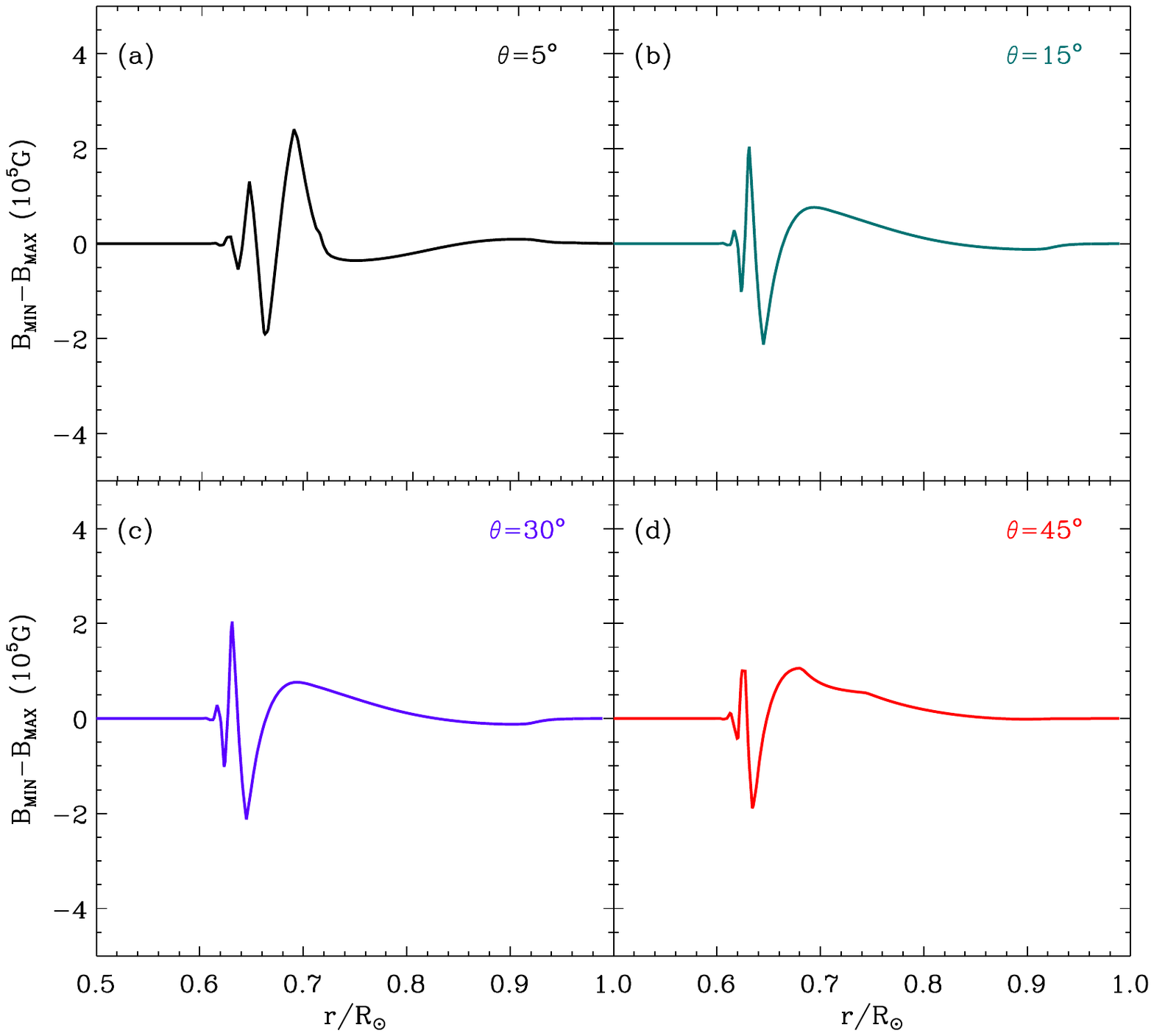}
  \caption{The changes in  magnetic  profiles between the minimum and the maximum of the SUYRA Standard Case at separated latitudes ($5^\circ$, $15^\circ$, $30^\circ$ and $45^\circ$).}
  \label{Fig:3}
  \end{figure}

  For a 1-D stellar evolution model, the time series of 2-D magnetic field has  been
  converted into 1-D data. As demonstrated in Figure~\ref{Fig:2}, the most magnetic activities
  appear at low and middle latitudes, hence we focus on a belt region from equator to
  $45^\circ$. The toroidal magnetic field distributed at nine latitude regions, i.e.,
  $5^\circ\pm2.5^\circ$, $10^\circ\pm2.5^\circ$, $15^\circ\pm2.5^\circ$, $20^\circ\pm2.5^\circ$, $25^\circ\pm2.5^\circ$, $30^\circ\pm2.5^\circ$,
  $35^\circ\pm2.5^\circ$, $40^\circ\pm2.5^\circ$ and $45^\circ\pm2.5^\circ$, are picked up and averaged individually as 1-D data. These 1-D data are assimilated in the 1-D solar model one at a time. Subsequently, we study the effects of magnetic field at one latitude region. Note that this is a simple and rough approximation and turns out to be the major limitation of this approach. Because the purpose of this work is to investigate the changes caused by magnetic field, we hence care the differences of magnetic strength rather than the absolute value. Figure~\ref{Fig:3} shows the radial distributions of the magnetic field difference between solar minimum and maximum. The four panels correspond to four latitudes., i.e., 5$^\circ$, 15$^\circ$, 30$^\circ$, and 45$^\circ$. As is shown that, the major changes are located below the base of convection zone, and no noticeable difference appears above 0.8$R_{\odot}$. This is because the SUYRA code does not consider a second dynamo near the surface. Since this work focuses on the helioseismic signals at the tachocline, the dynamo model is adequate.

\section{Structural Variations in the solar interior} \label{sec:models}
   The solar model we use in this paper is an established one obtained by \citet{bi11}. The model was calculated
  by the one-dimensional Yale Rotating Stellar Evolution Code \citep[YREC;][]{guether92,li03}.
  The OPAL equation of state tables EOS2005 \citep{rogers02}  and the OPAL high-temperature
  opacities \footnote{http://opalopacity.llnl.gov/new.html}
  GS98 \citep{grevesse98} supplemented by the low-temperature opacities \citep{ferguson05}  are
  adopted. The atmospheric model is constructed using the empirical Krishna-Swamy T-relation.
  Elements diffusion \citep{thoul94} is also taken into account.

  We use the solar variable model \citep{li03} to assimilate
  the magnetic field (1-D data) into the solar
  structural model.
  The magnetic field is described by $\chi$ and $\gamma$,
  which change the equilibrium of the model, and then we let YREC re-scale
  the solar model 20 times in order to build a new equilibrium model. Note that we have tested beforehand and 20 iterations are enough for the YREC code to restructure the solar model for all input magnetic field.
  When a solar model is resolved, we then assimilate the next magnetic field, re-scale the model for another equilibrium model.
  A series of solar models is finally obtained for a given time series of the magnetic field.

\subsection{ Magnetic impacts on the Interior Structures}

  The series of solar models  records the structural variations generated by
  the magnetic field in a solar cycle. Figures 4a and 4b show the changes of density and
  sound speed between the minimum and the maximum. As mentioned
  in Section 2.2, the  impacts of magnetic  field can change the stellar structure
  through the thermodynamic effect of magnetic pressure and magnetic energy.
  Because of the plasma-$\beta$ ($\beta=P/P_{\rm m})\gg 1$ in the deep interior,
  the magnetic field is certainly weak enough that it is only a small perturbation
  to the underlying structure. Although the local total pressure changes when the magnetic
  strength increases or decreases, the restoring time of the pressure and the
  temperature is short enough to keep little change in the numerical results.
  As a result, the density changes with the magnetic field, which brings variations
  in the sound speed, i.e., $c^{2}=\frac{\Gamma_{1}P}{\rho}$ where $\Gamma_{1}$
  is the first adiabatic exponent. Figures 4a-4b clearly show that the change in the square
  of sound speed is of the same order of magnitude as the density change.
  The differences in the global parameters between two extremum values are strongly
  consistent with the change in interior structure near the base of the convection zone.

  \begin{figure}
  \centering
  \includegraphics[width=100mm, angle=0]{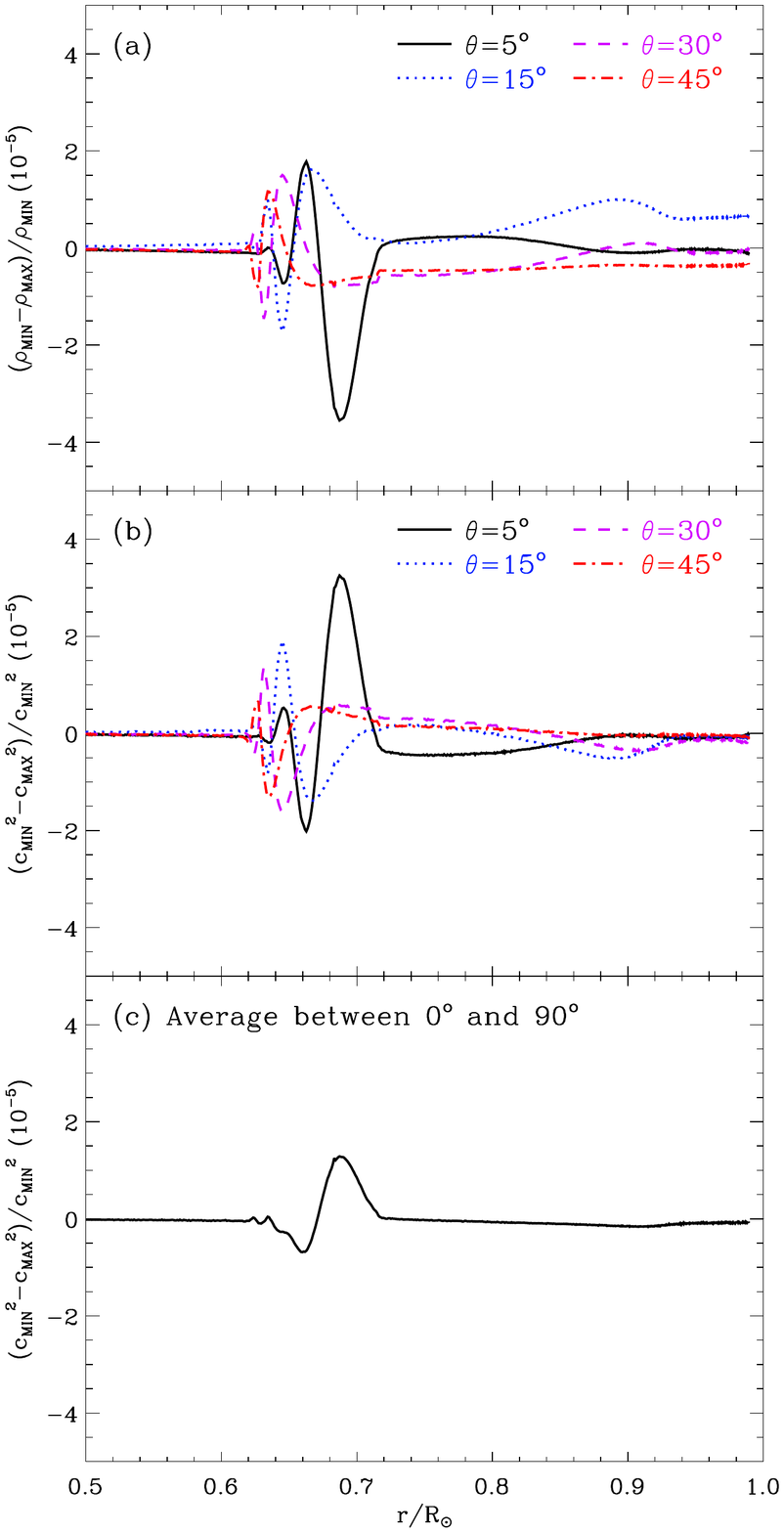}
  \caption{Calculated relative variations of density and sound speed
  between the minimum to the maximum as functions of radius. Average change
  in sound speed from $0^\circ$ to $90^\circ$.}
  \label{Fig:4}
  \end{figure}

  To compare the results of the change in sound speed with what has been obtained by
  helioseismology, we give the averaged variations of the sound speed between
  the minimum and the maximum, as shown in Figure 4c. The theoretical averaged changes in sound speed at
  all latitudes are included, and the average
  value is calculated with
  \begin{equation}
  \overline f  = \frac{\int^{90^\circ}_{\rm 0}f(\theta)\cos(\theta)d\theta}
  {\int^{90^\circ}_{\rm 0}\cos(\theta)d\theta},
  \end{equation}
  where $f$ is $(c^{2}_{\rm MIN}-c^{2}_{\rm MAX})/c^{2}_{\rm MIN}$.

  As seen in Figure 4c, the theoretical model displays significant variations of sound speed at
  the base of the convective zone. In addition, the 'S' shape of the sound-speed change is very similar to the helioseismic result.
  However, the largest relative deviation in the sound speed of
  $\delta c^{2}/c^{2}$ is only about 0.8-1.5$\times$ 10$^{-5}$, appearing at around
  $0.7R_{\odot}$, which is smaller than the helioseismic result
  (7.23$\pm $0.28$\times$ 10$^{-5}$) of \citet{baldner08} by an order of magnitude.
  Because of the simple treatment of magnetic field and the poor approximation of 2-D data, the 1-D solar model is apparently not sufficient. For instance, it does not includes the turbulent pressure which should be impacted by magnetic field. It should also be noted that the BL dynamo model ignores the comprehensive understanding of the dynamics of turbulence in the convection zone, which could lead to significant changes in the magnetic strength.
  Although there are a number of limitations in this framework, it is fair to use the solar model as a poor but consistent 'scale' to measure relative changes. For this reason, the agreement of the 'S' shape is still meaningful. This shape infers how the magnetic field varies its structure in a solar cycle and the structures at different time points are key constraints to the dynamo theories. The similar shapes found in above results hence infer that the BL model seems to provide a sensible dynamo process that fits the helioseismic findings.

\subsection{Solar Variable Models with Stronger Magnetic field}

\begin{figure}
  \centering
  \includegraphics[width=100mm, angle=0]{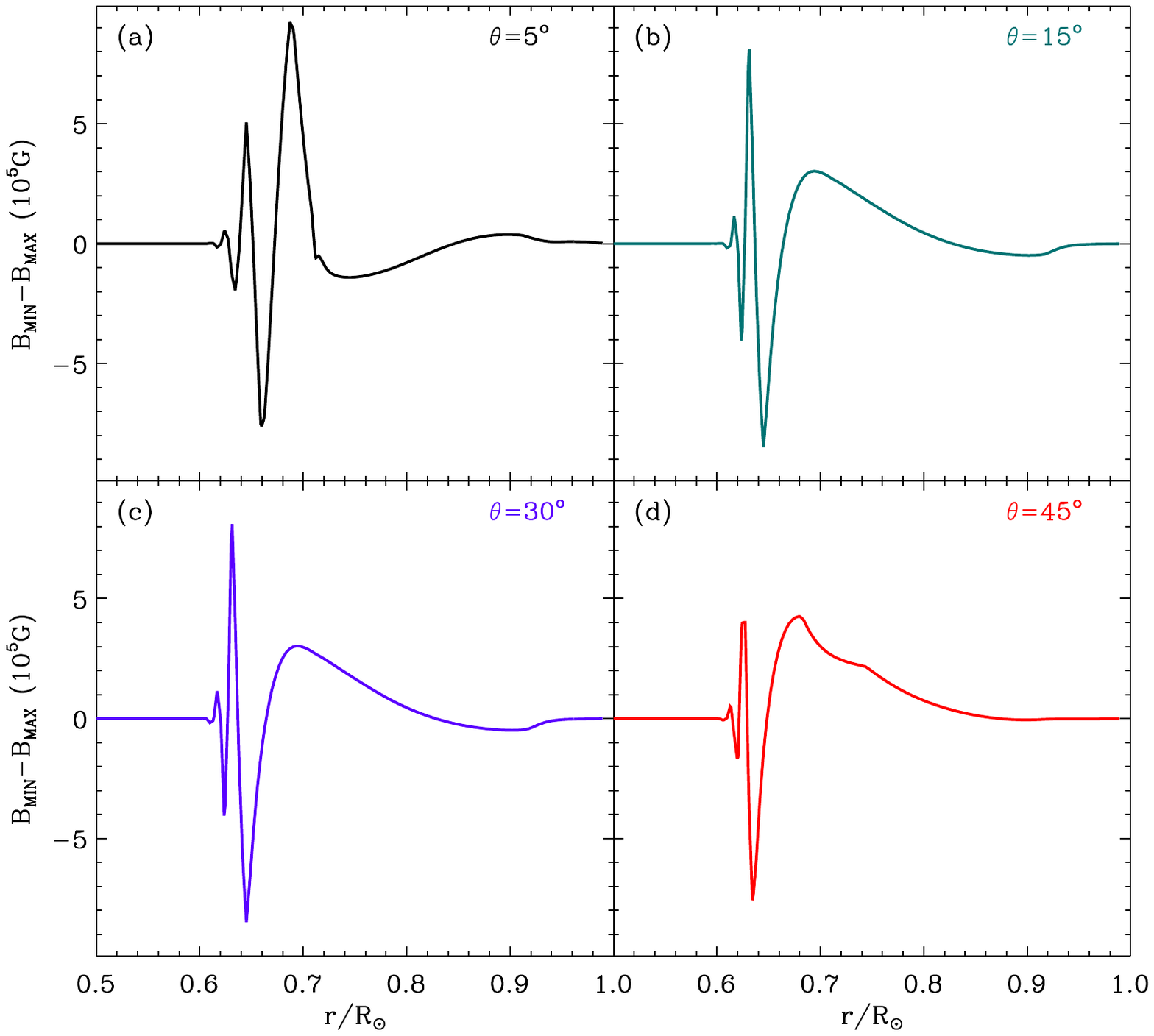}
  \caption{Same as Figure \ref{Fig:3} but for SUYRA Case 10.}
  \label{Fig:case7db}
  \end{figure}

\begin{figure}
  \centering
  \includegraphics[width=100mm, angle=0]{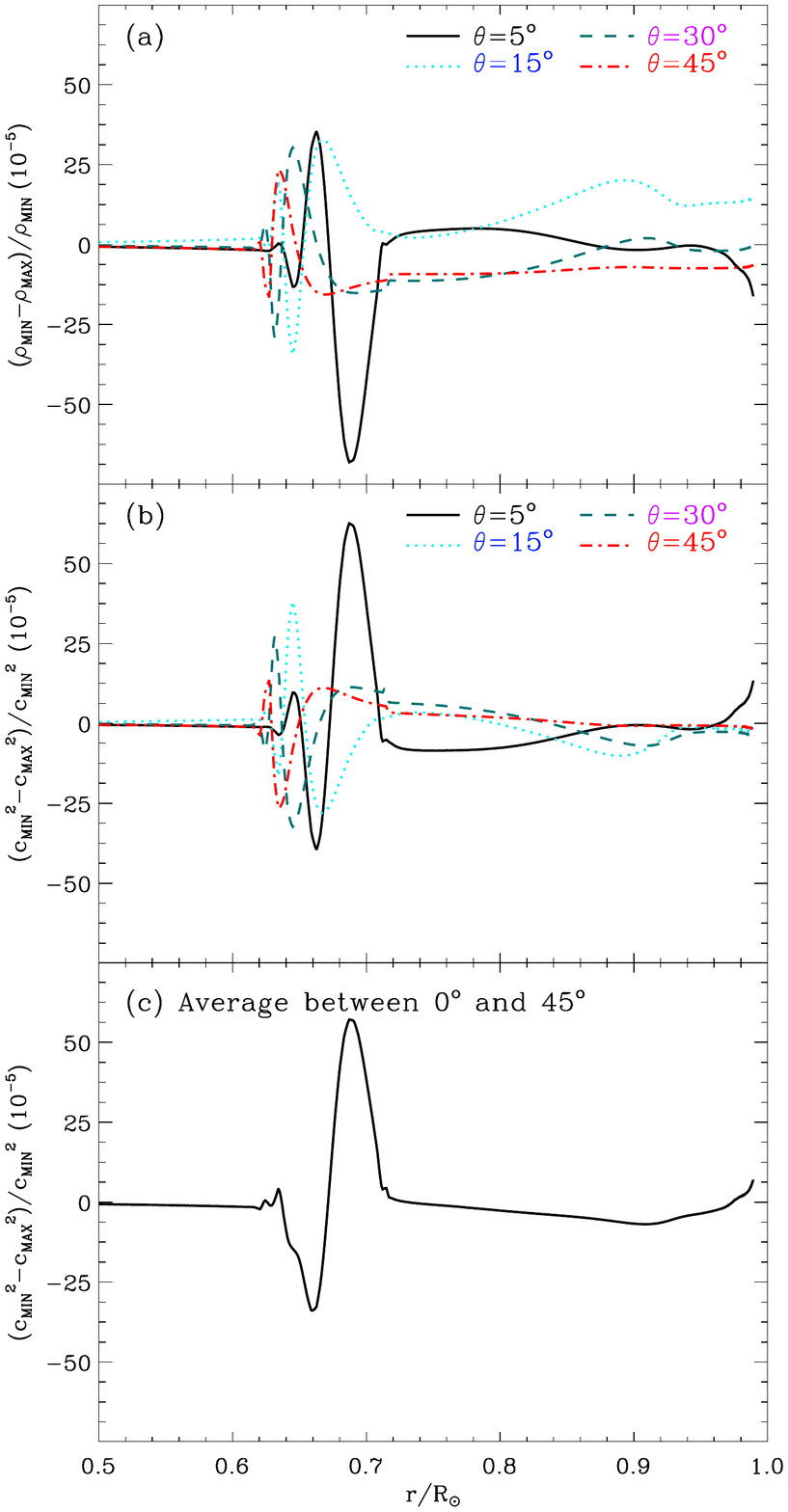}
  \caption{Same as Figure \ref{Fig:4} but for SUYRA Case 10.}
  \label{Fig:case7ss}
  \end{figure}

The above model has achieved a similar structural features of the change of sound speed. In this section, we adjust the three input parameters of the SUYRA code to generate stronger magnetic field that can cause similarly large changes as helioseismic results. We adjusted the critical field $B_{c}$, the base of convective zone $r$, and the fraction $f$ to modulate the magnetic strength. Note that a successful SUYRA model should also produce similar observed features including cyclic dipolar parity, butterfly diagram, and distribution of diffuse radial field at the surface. Rather than mentioning all solutions, here we only demonstrate the case (SUYRA Case 10 hereafter) which shows the best agreement with the helioseismic findings. The three adjusted parameters of SUYTA Case 10 are $B_{c} = 4\times10^5$G, $r$ = 0.71, and $f$ = 0.5. The variation of magnetic profiles of Case 10 is illustrated in Figure \ref{Fig:case7db}. The structural features are similar to the SUYRA Standard Case but the amplitude goes up to 6$\times10^5$G, which is twice of the Standard Case. Corresponding change in sound speed are shown in Figure \ref{Fig:case7ss}. Similar to the standard case, the average sound-speed change presents an 'S' shape but the absolute values are closed to helioseismic results.
The results indicate that our solar variable models require magnetic field on average of $\sim 4\times10^5$G to reproduce similar sound-speed changes in the tacholine. However, the strong toriodal field in the tachocline leads to unrealistically magnetic strength at the surface, which ranges from 100 to 1000G in the solar cycle and apparently too strong compared with the observations.

\section{Discussion and Conclusion}

Compared to \citet{li03}, the major differences are twofold. One is that the magnetic field to be incorporated into the solar model is more self-consistent since the BL type dynamo model was confirmed to be at the essence of solar cycle. However, the averaged magnetic field given by the dynamo model is too small to account for the changes in frequency
  above the latitude of $5^\circ$. This means that the kinematic modeling of solar cycle was rather an approximation procedure,
  which possibly ignored the effects of nonlinear interaction among magnetic field, convection and
  differential rotation on the dynamo. The other is that our results either suggest stronger field strength near surface layers,
  or indicate that the magnetic effects on the frequency is in a way different from the
  assumptions in our solar models.

 We find that the changes in sound speed near the base
  of the convection zone are strongly consistent with the change in the interior structure
  in the solar cycle, which are roughly close to the obtained values. Moreover,
  it has been shown that the change in frequency is tightly correlated with
  the spatial distribution of the surface magnetic field. The significant shifts have not found
  in our models, implying that the shifts cannot be purely explained by structural changes due
  to cycle field generated by the BL type dynamo model.

  In this work, we developed a 1-D solar model to study the effects of solar dynamo on the solar internal structures. Although there are several limitations in the current framework, it offers a tool to use helioseismic findings to constrain the the profile and the strength of the internal magnetic field.
  For further investigation of the relevant solar cyclic variations and
  stellar cycles to more stars, we should extend studies of the interior and surface dynamical processes,
  including the roles of turbulence, the flux emergence, the nonlinearities and so on.

\begin{acknowledgements}
  Sincerely, we thank the developers of Solar Dynamo code SURYA, Arnab
  Rai Choudhuri and his collaborators from Department of Physics of
  Indian Institute of Science, for their fundamental works. This work is
  supported by the Joint Research Fund in Astronomy (U1631236 and U2031203)
  under cooperative agreement between the National Natural Science
  Foundation of China (NSFC) and Chinese Academy of Sciences (CAS),
  and NSFC 11522325. This work has also received funding from the European Research Council (ERC) under the European Union’s Horizon 2020 research and innovation programme (CartographY GA. 804752).
\end{acknowledgements}

\bibliographystyle{raa}

\bibliography{ref}

\begin{thebibliography}{44}
\providecommand\natexlab[1]{#1}
\providecommand\JournalTitle[1]{#1}

\bibitem[{Babcock}(1961)]{babcock61}
{Babcock}, H.~W. 1961, \apj, 133, 572

\bibitem[{Baldner} {et~al.}(2009)]{baldner09}
{Baldner}, C.~S., {Antia}, H.~M., {Basu}, S., \& {Larson}, T.~P. 2009, \apj,
  705, 1704

\bibitem[{Baldner} \& {Basu}(2008)]{baldner08}
{Baldner}, C.~S., \& {Basu}, S. 2008, \apj, 686, 1349

\bibitem[{Basu} \& {Schou}(2000)]{basu00}
{Basu}, S., \& {Schou}, J. 2000, \solphys, 192, 481

\bibitem[{Bi} {et~al.}(2011)]{bi11}
{Bi}, S.~L., {Li}, T.~D., {Li}, L.~H., \& {Yang}, W.~M. 2011, \apjl, 731, L42

\bibitem[{Broomhall} {et~al.}(2009)]{broomhall09}
{Broomhall}, A.~M., {Chaplin}, W.~J., {Elsworth}, Y., {Fletcher}, S.~T., \&
  {New}, R. 2009, \apjl, 700, L162

\bibitem[{Cameron} {et~al.}(2010)]{cameron10}
{Cameron}, R.~H., {Jiang}, J., {Schmitt}, D., \& {Sch{\"u}ssler}, M. 2010,
  \apj, 719, 264

\bibitem[{Chaplin} {et~al.}(2007)]{chaplin07}
{Chaplin}, W.~J., {Elsworth}, Y., {Miller}, B.~A., {Verner}, G.~A., \& {New},
  R. 2007, \apj, 659, 1749

\bibitem[{Charbonneau}(2010)]{charbonneau}
{Charbonneau}, P. 2010, Living Reviews in Solar Physics, 7, 3

\bibitem[{Charbonneau}(2014)]{charbonneau14}
{Charbonneau}, P. 2014, \araa, 52, 251

\bibitem[{Chatterjee} {et~al.}(2004)]{chatterjee04}
{Chatterjee}, P., {Nandy}, D., \& {Choudhuri}, A.~R. 2004, \aap, 427, 1019

\bibitem[{Chou} \& {Serebryanskiy}(2005)]{chou05}
{Chou}, D.-Y., \& {Serebryanskiy}, A. 2005, \apj, 624, 420

\bibitem[{Choudhuri}(2017)]{Choudhuri2017}
{Choudhuri}, A.~R. 2017, Science China Physics, Mechanics, and Astronomy, 60,
  19601

\bibitem[{Choudhuri}(2020)]{Choudhuri2020}
{Choudhuri}, A.~R. 2020, arXiv e-prints, arXiv:2008.09347

\bibitem[{Christensen-Dalsgaard} {et~al.}(1996)]{christensen-Dalsgaard}
{Christensen-Dalsgaard}, J., {Dappen}, W., {Ajukov}, S.~V., {et~al.} 1996,
  Science, 272, 1286

\bibitem[{Denissenkov} \& {Pinsonneault}(2007)]{denissenkov07}
{Denissenkov}, P.~A., \& {Pinsonneault}, M. 2007, \apj, 655, 1157

\bibitem[{Eggenberger} {et~al.}(2008)]{eggenberger08}
{Eggenberger}, P., {Meynet}, G., {Maeder}, A., {et~al.} 2008, \apss, 316, 43

\bibitem[{Elsworth} {et~al.}(1990)]{elsworth90}
{Elsworth}, Y., {Howe}, R., {Isaak}, G.~R., {McLeod}, C.~P., \& {New}, R. 1990,
  \nat, 345, 322

\bibitem[{Ferguson} {et~al.}(2005)]{ferguson05}
{Ferguson}, J.~W., {Alexander}, D.~R., {Allard}, F., {et~al.} 2005, \apj, 623,
  585

\bibitem[{Grevesse} \& {Sauval}(1998)]{grevesse98}
{Grevesse}, N., \& {Sauval}, A.~J. 1998, \ssr, 85, 161

\bibitem[{Guenther} {et~al.}(1992)]{guether92}
{Guenther}, D.~B., {Demarque}, P., {Kim}, Y.~C., \& {Pinsonneault}, M.~H. 1992,
  \apj, 387, 372

\bibitem[{Hotta} {et~al.}(2016)]{hotta2016}
{Hotta}, H., {Rempel}, M., \& {Yokoyama}, T. 2016, Science, 351, 1427

\bibitem[{Howe} {et~al.}(2018)]{howe18}
{Howe}, R., {Chaplin}, W.~J., {Davies}, G.~R., {et~al.} 2018, \mnras, 480, L79

\bibitem[{Howe} {et~al.}(2000)]{howe00}
{Howe}, R., {Christensen-Dalsgaard}, J., {Hill}, F., {et~al.} 2000, Science,
  287, 2456

\bibitem[{Howe} {et~al.}(2002)]{howe02}
{Howe}, R., {Komm}, R.~W., \& {Hill}, F. 2002, \apj, 580, 1172

\bibitem[{Jain} {et~al.}(2009)]{jain09}
{Jain}, K., {Tripathy}, S.~C., \& {Hill}, F. 2009, \apj, 695, 1567

\bibitem[{Jiang} {et~al.}(2013)]{jiang13}
{Jiang}, J., {Cameron}, R.~H., {Schmitt}, D., \& {I{\textcommabelow s}{\i}k},
  E. 2013, \aap, 553, A128

\bibitem[{Jiang} {et~al.}(2007)]{jiang07}
{Jiang}, J., {Chatterjee}, P., \& {Choudhuri}, A.~R. 2007, \mnras, 381, 1527

\bibitem[{Kosovichev}(1996)]{kosovichev}
{Kosovichev}, A.~G. 1996, \apjl, 469, L61

\bibitem[{Leighton}(1964)]{leighton64}
{Leighton}, R.~B. 1964, \apj, 140, 1547

\bibitem[{Li} {et~al.}(2003)]{li03}
{Li}, L.~H., {Basu}, S., {Sofia}, S., {et~al.} 2003, \apj, 591, 1267

\bibitem[{Li} \& {Sofia}(2001)]{li01}
{Li}, L.~H., \& {Sofia}, S. 2001, \apj, 549, 1204

\bibitem[{Liang} \& {Chou}(2015)]{liang15}
{Liang}, Z.-C., \& {Chou}, D.-Y. 2015, \apj, 809, 150

\bibitem[{Libbrecht} \& {Woodard}(1990)]{libbrecht90}
{Libbrecht}, K.~G., \& {Woodard}, M.~F. 1990, \nat, 345, 779

\bibitem[{Lydon} \& {Sofia}(1995)]{lydon95}
{Lydon}, T.~J., \& {Sofia}, S. 1995, \apjs, 101, 357

\bibitem[{Meynet} \& {Maeder}(1997)]{meynet97}
{Meynet}, G., \& {Maeder}, A. 1997, \aap, 321, 465

\bibitem[{Nandy} \& {Choudhuri}(2002)]{nandy02}
{Nandy}, D., \& {Choudhuri}, A.~R. 2002, Science, 296, 1671

\bibitem[{Rogers} \& {Nayfonov}(2002)]{rogers02}
{Rogers}, F.~J., \& {Nayfonov}, A. 2002, \apj, 576, 1064

\bibitem[{Schou} {et~al.}(1998)]{schou98}
{Schou}, J., {Antia}, H.~M., {Basu}, S., {et~al.} 1998, \apj, 505, 390

\bibitem[{Serebryanskiy} \& {Chou}(2005)]{serebryanskiy05}
{Serebryanskiy}, A., \& {Chou}, D.-Y. 2005, \apj, 633, 1187

\bibitem[{Strugarek} {et~al.}(2017)]{strugarek2017}
{Strugarek}, A., {Beaudoin}, P., {Charbonneau}, P., {Brun}, A.~S., \& {do
  Nascimento}, J.~D. 2017, Science, 357, 185

\bibitem[{Thoul} {et~al.}(1994)]{thoul94}
{Thoul}, A.~A., {Bahcall}, J.~N., \& {Loeb}, A. 1994, \apj, 421, 828

\bibitem[{Tripathy} {et~al.}(2015)]{tripathy15}
{Tripathy}, S.~C., {Jain}, K., \& {Hill}, F. 2015, \apj, 812, 20

\bibitem[{Woodard} \& {Noyes}(1985)]{woodard}
{Woodard}, M.~F., \& {Noyes}, R.~W. 1985, \nat, 318, 449

\end{thebibliography}

\end{document}